# Application of Orthogonal Defect Classification for Software Reliability Analysis

**Edward Chen[a], Han Bao[b], Tate Shorthill[c], Carl Elks[d], Athira Varma Jayakumar[e], Nam Dinh[f]**

[a]North Carolina State University, 2500 Stinson Dr., Raleigh, NC, 27607, USA, echen2@ncsu.edu
[b]Idaho National Laboratory, P.O. Box 1625, MS 3860, Idaho Falls, ID 83415, USA, han.bao@inl.gov
[c]University of Pittsburgh, 3700 O'Hara Street, Pittsburgh, PA 15261, USA, ths60@pitt.edu
[d]Virginia Commonwealth University, Richmond, VA, USA, crelks@vcu.edu
[e]Virginia Commonwealth University, Richmond, VA, USA, jayakumarav@vcu.edu
[f]North Carolina State University, 2500 Stinson Dr., Raleigh, NC, 27607, USA, ntdinh@ncsu.edu

**Abstract:** The modernization of existing and new nuclear power plants with digital instrumentation and control systems (DI&C) is a recent and highly trending topic. However, there lacks strong consensus on best-estimate reliability methodologies by both the United States (U.S.) Nuclear Regulatory Commission (NRC) and the industry. This has resulted in hesitation for further modernization projects until a more unified methodology is realized. In this work, we develop an approach called Orthogonal-defect Classification for Assessing Software Reliability (ORCAS) to quantify probabilities of various software failure modes in a DI&C system. The method utilizes accepted industry methodologies for software quality assurance that are also verified by experimental or mathematical formulations. In essence, the approach combines a semantic failure classification model with a reliability growth model to predict (and quantify) the potential failure modes of a DI&C software system. The semantic classification model is used to address the question: How do latent defects in software contribute to different software failure root causes? The use of reliability growth models is then used to address the question: Given the connection between latent defects and software failure root causes, how can we quantify the reliability of the software? A case study was conducted on a representative I&C platform (ChibiOS) running a smart sensor acquisition software developed by Virginia Commonwealth University (VCU). The testing and evidence collection guidance in ORCAS was applied, and defects were uncovered in the software. Qualitative evidence, such as condition coverage, was used to gauge the completeness and trustworthiness of the assessment while quantitative evidence was used to determine the software failure probabilities. The reliability of the software was then estimated and compared to existing operational data of the sensor device. It is demonstrated that by using ORCAS, a semantic reasoning framework can be developed to justify if the software is reliable (or unreliable) while still leveraging the strength of the existing methods.

## 1. INTRODUCTION

In recent years, there has been considerable effort to modernize existing and new nuclear power plants with digital instrumentation and control systems (DI&C). However, there has also been a considerable concern both by industry and regulatory bodies on the risk and consequence analysis of these systems. The lack of a strong consensus on best-estimate methodologies by both the United States (U.S.) Nuclear Regulatory Commission (NRC) and the industry [1] has led to a hesitation for further modernization projects. In branch technical position (BTP) 7-19 [2], the NRC has also cited concerns that DI&C systems can be vulnerable to common cause failures (CCFs) because of software errors in logic or implementation that could reduce defense-in-depth capability in existing hardware redundant architectures. Such software errors can manifest due to inadequacies in either the design requirements specifications or the implementation of the design.

While many new methods have been proposed to identify potential software events, such as Systems Theoretic Process Analysis (STPA) [3], Hazard and Consequence Analysis for Digital Systems (HAZCADS) [4], etc., these methods are focused on the qualitative identification of failure modes in a fault tree with very little guidance on direct quantification. Typically, software failure modes are identified by potential unsafe control actions (UCA) made by the system [3]. The UCAs can lead to stakeholder losses and are traceable to a particular system level event which can be integrated as basic events within fault trees [5]. Supplemental assessments to HAZCADS, such as the Digital Reliability Assessment Methodology (DRAM) [6], have been used to address the risk of UCA by assigning Risk Reduction Targets (RRTs) and Control Methods (CMs) to bound the risk of software basic events. The risk to the system is determined by the Safety Integrity Level (SIL) [7] intended for the device. However, RRTs and CMs are qualitative methods at risk mitigation. Specifically, the approach helps identify failure mechanisms and pathways that can lead to UCAs and methods to address them. CMs listed in DRAM are scored qualitatively based on implemented type and effectiveness by expert experience ad belief. However, the improvement in reliability by CMs and RRTs are difficult to quantify. Nonetheless, DRAM provides a useful qualitative support identification framework for design activities.

Aside from bounding estimate methods, other more direct risk and reliability quantification methods include software reliability growth models (SRGMs) [8]. These well-known methods attempt to predict the anticipated reliability of the software through historical failure data and has historically good generalization across multiple industries. However, conventional use of SRGMs are to measure the wholistic reliability of software and rather than specific subsystems (due to the lack of failure data). In cases where failure data is limited, which is especially true for safety critical systems, the uncertainty in the SRGMs can render predictions meaningless [9]. Our research thus aims to provide more acceptable risk and reliability information on DI&C systems without losing specificity or generalization capability. Furthermore, we attempt to collect qualitative evidence to support reliability conclusions.

In this work, we present the idea of Orthogonal-defect Classification for Assessing Software Reliability (ORCAS) to formalize and provide actionable evidence for the reliability quantification of DI&C system. The method utilizes accepted industry testing methodologies for software quality assurance that have also been verified by experimental or mathematical formulations. A pseudo-exhaustive testing [10] approach is adopted to reduce the cost of testing while maintaining a similar level of coverage confidence. In essence, ORCAS combines a semantic failure classification model with a reliability growth model to predict (and quantify) potential failure modes of a DI&C software system. The semantic classification model is used to address the question: How do latent defects in software translate to different software failure modes? For example, suppose a defect was discovered that fails to check a variable's contents before using it in an equation. The direct impact of this defect on the software is difficult to gauge; it may have no impact or have serious consequences in calculation. By translating discovered defects into defined independent categorical types, the impact to software reliability can be generalized and modeled. Here, the use of reliability growth models is more applicable: Given that we know how latent defects contribute to different software failure modes, the risk (or reliability) of the DI&C system can be predicted. In addition, derived qualitative evidence from the ORCAS methodology is used to determine the confidence in our assessment and whether the results are trustworthy.

A case study was conducted on a representative software platform (ChibiOS) [11] running a sensor acquisition software developed by Virginia Commonwealth University (VCU) [12]. The testing and evidence collection guidance in ORCAS was applied. Defects were uncovered in the software. Qualitative evidence, such as condition coverage, was used to gauge the completeness and trustworthiness of the assessment while quantitative evidence was used to determine software failure probabilities. The reliability of the software was then estimated and compared to existing operational data of the sensor device. It is demonstrated that by using ORCAS, a semantic reasoning framework can be developed to justify software reliability (or unreliability) while still leveraging the strength of existing methods.

## 2. THEORETIC BACKGROUND

Before presenting the methodology, the theory behind this work is first discussed. The most important theory being that 'failures' in software are ill-defined. In STPA, the authors treat software 'failures' as misbehaviors or unintended consequences due to inadequate (but deliberate) specification of constraints and requirements of the system [3]. In this respect, the software never truly 'fails' but rather performs actions that are undesirable while still conforming to existing requirements. Hence, the development of

UCA in STPA. In this work, the same ideology is utilized, where the different categories of UCA are referred to as software failure modes. In brief, the failure modes are: (1) control action is missing when needed (UCA-A); (2) a control action is provided when not needed (UCA-B); (3) a control action is provided but too early, too late, or out of order (UCA-C); and (4) a control action is stopped too soon or applied too long (UCA-D). However, it is unclear how latent defects in the software translate to each failure mode. For example, missing conditional statement are a common software defect but their relationships to a particular UCA is uncertain. In ORCAS, this relationship is refined with Orthogonal Defect Classification (ODC).

The second theory used in this work is ODC [13]. In ODC, discovered defects are the root causes of software failure modes. However, the cause-and-effect relationship between these root causes and failure mode is not always clear unless explicit cases can be demonstrated. Here ODC acts as a semantic classification bridge between these cause-and-effect relationships by sorting defects into generalized groups. Specifically, by grouping defects with shared characteristics, the characteristics of the failure caused by individual defects can also be generalized. For example, if checking defects are detected, it suggests inadequacy in data or condition verification in the source code. This can then be traced to higher UCA-A probability based on derived correlations. The exact correlations are discussed in the methods section. In brief, the defect groups are function, assignment, algorithm, checking, interface, relationship, and timing [14]. These groups are assumed to be independent and mutually exclusive from each other and cover all known to-date potential defects in the software. In addition, a defect can only be assigned one defect class, but may cause multiple UCAs. In ORCAS, the defect classes are used as qualitative evidence to assess software reliability and causality to software failure modes.

Another application of ODC is the identification of the necessary environmental and input conditions required to uncover or detect defects. These conditions are known as triggers and can also be used to assess when all conditions have been considered. From a conventional perspective, testing conditions as triggers are difficult to measure and compare with each other as tests are not equivalent. Here, ODC also can be used to semantically categorize triggers that are needed for comprehensive software testing. In brief, the groups for implementation triggers are simple and complex path; test coverage, variation, sequence, and interaction; volume/stress; recovery; configuration; startup/restart; and normal mode [14]. These trigger groups are extensive and cover most relevant scenarios, but it is not a complete list. For instance, they do not explicitly consider cybersecurity vulnerabilities as a condition for defects. Nonetheless, from a development perspective, the trigger groups represent an adequate scope for required testing conditions and have been used extensively by the industry [15]. In ORCAS, triggers are used as qualitative evidence to assess the completeness of the testing effort.

This work is part of the Light Water Reactor Sustainability (LWRS) project, "DI&C Risk Assessment". The object of this project is to provide effective quantitative and qualitative measurement tools to gauge the risk and reliability of modernization projects in existing and novel nuclear reactors [16] [17] [18]. In the LWRS-developed framework for DI&C risk assessment, ORCAS is developed and used to quantify the probability of STPA-identified UCAs [19] [20] [21].

## 3. METHODOLOGY

The basis for ORCAS is to use pseudo-exhaustive test-based approaches [10] to generate a historical failure database. Defects that are detected and removed are classified based on ODC theory. Each defect class is then modeled to quantify probabilities of different software failure modes. Qualitative evidence collected throughout this process is then used to gauge how complete and confident we are in the assessment. The overall workflow of the method (and meaningful extensions) can be seen in Figure. 1. Items in the dashed box are all elements pertaining to the ORCAS methodologies. In general, the outputs of ORCAS are the software failure mode probabilities and confidence in the assessment. The qualitative evidence derived from ORCAS include the requirement traceability matrix (RTM), trigger coverage assessment (TCA), structural path coverage, and reliability modeling stability. The quantitative evidence include the defect reports used to determine failure probability.

## 3.1 Stages of ORCAS

It is important to note that the assessment of software reliability should be continuously evolving with the implementation and design of the system. For instance, the target and scope of the analysis may not be fully known at the start of the assessment due to inherent complexity of the software system. Missed relevant items will require returning to the prior stage for further refinement. For example, when a defect is detected and classified in stage 3, it is expected to be repaired before the software is deployed. This will require returning to stage 2 for defect removal activities. Imperfect knowledge and discovery at any individual stage suggests that revisiting a prior stage is anticipated.

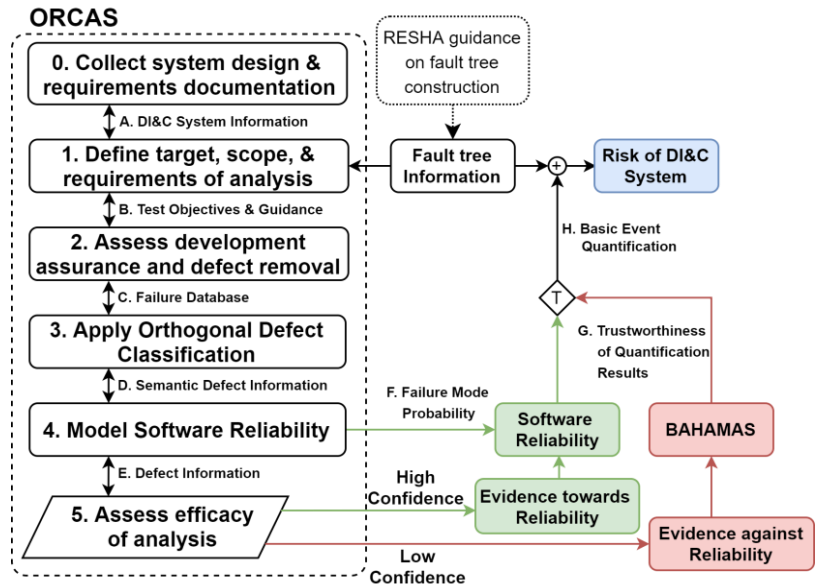

Figure. 1 Overall workflow of ORCAS.

In stage 0, the relevant information and details pertaining to the system are collected. This step is assumed to occur in any assessment and thus not described in detail. The information that can be collected at this stage can include formal documentation (i.e., IEEE 29148, IEEE 829, IEC 61508, etc.), defect and anomaly reports, design, and requirements specifications. Exact documents are not specified; however, information pertaining to the functional and non-functional requirements, implementation design, and test verification and validation are required for further stages.

In stage 1, the scope and testing adequacy requirements are defined. Here the desired modules and functions of the system are outlined and what type of testing is required. As software can be incredibly complex, initially assigning the scope to the entire system can be overwhelming and uninformative. Rather, it is advised that a fault tree approach is adopted to assess exactly what the stakeholders are concerned about (via top events) and how it may impact operational goals. The fault tree provides structure, but also linear relationships between software failures to loss events. HAZCADS can be one method to develop the qualitative fault tree; however, in this work, the REdundancy-guided Systems-theoretic Hazard Analysis (RESHA) [22] [23]is utilized to deconstruct the DI&C systems. In RESHA, failures in the control systems are based on physical separation of components and devices to emphasize the focus on control processes. For instance, an analog to digital converter integrated circuit is a physical component where its hardware and software failures can be modeled in an integrated fault tree. In addition, recent work in RESHA also introduces unsafe information flows (UIFs) as a software failure mechanism. UIFs are relevant for information-based, control-absent systems such as monitoring systems. In essence, UIFs mimic UCAs in failure categories but deals with information/feedback dependencies. The failure modes are: (1) failure to provide feedback when needed (UIF-A); (2) providing feedback when not needed (UIF-B); (3) the feedback that is provided comes too early, too late, or out of sync/sequence (UIF-C); and (4) the feedback value is low, high, not-a-number (NaN), or infinity (Inf) (UIF-D). Further details can be found in [17]. Lastly, RESHA fault tree construction follows an STPA top-down systems theoretic approach. This can help constrain the size of the tree to only basic failure events relevant to the stakeholders. Once the relevant aspects for analysis are identified, the target software can be assessed based on testing completeness (adequacy) and satisfaction of test formulation requirements. The purpose of these testing adequacy requirements is to determine what tests should have been implemented for verification and how complete the testing is. In this work, the adequacy of requirements is outlined in a three-level test suite hierarchy, as shown in Figure 2. Each level specifies defect triggers that need to be considered during test implementation and activities that should be conducted to address the testing adequacy. Recommended methodologies, such as T-way combinatorial testing [24], modified condition decision coverage (MCDC) [25], boundary value analysis (BVA) [26], and equivalence partitioning (EP) [26] have been proven to be experimentally effective and are recommended to complete each activity. However, quantitative test metrics alone are not adequate at predicting software reliability, as discovered in [27]. Hence, the need for qualitative evidence to assess trustworthiness and completeness in the assessment.

In stage 2, the testing adequacy requirements identified in the previous stage are compared to the testing efforts conducted by the development team. For activities involving T-way combinatorial testing, BVA, and EP, the specific range, variation, edges cases etc. of the parameters were identified in the previous stage. These need to be traceable to test cases implemented by the developer. For path analysis, tests should exist that consider different path conditions. For the RTM, both functional and non-functional requirements of the software should be traced to associated test cases that demonstrate conformance. The completeness of the RTM, TCA, and structural coverage is used as qualitative evidence for testing completeness. These metrics also serve to identify areas requiring further testing. For instance, if function variation was not considered during testing, the associated tests can be implemented to satisfy this metric. The defects from testing are collected from two sources. The first source are the existing defect reports during the development process. The second source is the additional defect removal process, which is initiated due to inadequate test coverage.

| Level 3 Testing (User) | | |
|---|---|---|
| Testing Activities | Defect Triggers Covered | Recommended Methods |
| System Test | Software Configuration | T-way Parameter Testing<br>Bounday Value Analysis<br>Equivalence Partitioning |
| System Test | Workload/Stress<br>Recovery/Exception<br>Startup/Restart<br>Normal Mode | Requirements Traceability Matrix |

| Level 2 Testing (Subsystem) | | |
|---|---|---|
| Testing Activities | Defect Triggers Covered | Recommended Methods |
| Unit Test | Simple Path<br>Complex Path | Modified Condition Decision Coverage |
| Function Test | Function Coverage<br>Function Variation<br>Function Squence<br>Function Interaction | T-way Parameter Testing<br>Boundary Value Asessment<br>Equivalence Partitioning |

| Level 1 Testing (Component) | | |
|---|---|---|
| Testing Activities | Defect Triggers Covered | Recommended Methods |
| Unit Test | Simple Path<br>Complex Path | Modified Condition Decision Coverage |
| Function Test | Function Coverage<br>Function Variation<br>Function Squence<br>Function Interaction | T-way Parameter Testing<br>Boundary Value Asessment<br>Equivalence Partitioning |

Figure 2. Three-tier software testing requirements with recommended activities and methods.

In stage 3, the defect reports are collected and categorized based on ODC theory [28]. An analysis of defects involves understanding what went wrong and how it was resolved. Importantly, defects are classified based on shared characteristics of the resolution or solution. It should be noted that if widely different solutions exist for the same problem, it may be an indication of inadequate requirements and constraints specification for the problem. After the classification of defects, the defect reports are assigned to specific software failure modes based on data-driven causality relationships.

The relationship between defect class and software failure mode is based on quantitative data collected from various open-source Github repositories namely, the MongoDB, Cassandra, Apache HBase, Zephyr, and OpenPilot repositories. The first three are all NoSQL database management systems; data can be found at [29]. Zephyr is a scalable small-footprint real-time operating system while OpenPilot is a semi-autonomous driving system. For each issue report in the databases, the defect class was determined first, followed by how it impacted the software. The impact to the users/stakeholders was generalized into the four UCA/UIF software failure modes. The contribution by each class to specific UCA/UIF then was counted and used to determine conditional probabilities (e.g., $P(\text{UCA-A}|\text{Function defect})$). In this work, 402 defect reports were used in total to form the conditional probabilities. In Figure 3, the various conditional probabilities with UCA/UIF can be seen. Darker color indicates greater correlation between defect class and UCA/UIF failure mode. For example, suppose a new assignment defect was detected in the software. Based on historical data, it has a 66.7% chance to cause a UCA/UIF-B under the worst environmental conditions. However, it is not as useful to discuss discovered defects, as they will have been removed/repaired. Therefore, the probability of remaining undetected defects by class is desirable.

|  | UCA/UIF-A | UCA/UIF-B | UCA/UIF-C | UCA/UIF-D |
|---|---|---|---|---|
| Algorithm Defects | 0.320 | 0.140 | 0.350 | 0.190 |
| Assignment Defects | 0.288 | 0.667 | 0.045 | 0.000 |
| Checking Defects | 0.360 | 0.244 | 0.256 | 0.140 |
| Function Defects | 0.389 | 0.222 | 0.241 | 0.148 |
| Interface Defects | 0.347 | 0.533 | 0.080 | 0.040 |
| Timing Defects | 0.190 | 0.048 | 0.524 | 0.238 |

Figure 3. Causality of software failure modes (UCA/UIFs) and orthogonal defect classes

In stage 4 of ORCAS, the probability of specific defect classes are determined. The methodology behind reliability growth models is not discussed in detail as extensive literature already exists. However, for ODC class specific SRGM, the author recommends Ref. [28]. In essence, each defect that is identified and classified requires the time or effort during testing. There are two important outcomes from this stage. The first is the predicted failure probability of each defect class and the second is the usefulness of the model. The second outcome is highly dependent on the amount of failure data required to generate reasonably accurate predictions on the remaining number of defects. Stability is used as a qualitative attribute to gauge usefulness. Generally, in stable models, the total number of predicted defects should not vary significantly from week to week. If variations are large, the resulting predictions also have a large variation thus rendering the model useless. In Ref. [30], this was the issue for their safety critical DI&C system as there was insufficient failure data for convergence. A 10% maximum variation allowance is recommended and shown to be effective [31]. This value also informs developers when sufficient failure data has been collected and testing can be stopped.

It is anticipated that reliability growth modeling will be difficult and not always applicable in every software development life cycle (SDLC). The measurement of effort can be difficult to accurately determine due to a range of factors. Development groups may batch repair defects together or the software development may be proprietary, making SRGM parameter estimation difficult. In such cases, a bounded failure probability estimation approach can be adopted. In bounded estimation, it is assumed that most defects have already been removed. Following reliability growth theory, further testing effort only reduces the probability of a defect existing; thus, the failure probability will only decrease. If such an approach is used, it is recommended only after the SDLC where testing has been conducted. The probability is determined approximately by counting the number of defects by class and dividing by the total testing effort. For failure-on-demand, the test effort is the number of tests conducted. For continuous failure probability, the test effort is the product of the test duration by the number of tests. The output is the failure probability of each defect class (e.g., $P(\text{Defect})$).

Finally, the probability of specific UCA/UIF classes can be determined by multiplying the conditional probabilities of each class with an individual class probability, as shown in Eqn.(1. Recall the conditional probability was pre-determined through the historical defect data. Only the second term, $P(\text{Defect})$ needs to be determined by the users. The result is the probability of each UCA/UIF mode (Eqn.1). The total software failure probability can be found through the sum of all UCA/UIF modes.

$$P(\text{UCA}_X) = P(\text{UCA}_X|\text{Alg.})P(\text{Alg.}) + \cdots + P(\text{UCA}_X|\text{Timing.})P(\text{Timing.}) \quad (1)$$

The last stage of ORCAS is the qualification of the software development process. Recall that the qualitative information derived from the method include the RTM, the TCA, the structural path coverage, and the stability of the reliability modeling. The developers and users can assess the completeness of the testing effort by reviewing how complete each qualitative factor is and which areas need further refinement. For instance, the RTM informs the developers whether each requirement was tested, while the TCA informs the developers that every scenario considered. The developers can then return to those software sections and conduct further testing.

### 3.1 Assumptions and Limitations

This method has several assumptions and limitations. The most concerning limitation is the use of reliability growth modeling in highly critical systems. In Ref. [30], insufficient software failure data was cited as one reason why this method is infeasible. When operational failure data is limited, the model is oversensitive, and the predictions have large uncertainties. While this may not be the case for all software, the lack of failure data is a highly relevant and limiting scenario. ORCAS only partially addresses this issue. When testing completeness is sufficient, but failure data is insufficient (i.e., all triggers are considered but no defects were discovered), ORCAS defers to a different methodology, which is known as Bayesian and Human-reliability-analysis Aided Method for the reliability Analysis of Software (BAHAMAS) [32]. The qualitative evidence derived from ORCAS can also be used to support BAHAMAS. This is seen in Figure. 1, where low assessment confidence leads to BAHAMAS.

The second major assumption is that causality between defect classes and UCAs/UIFs do not differ significantly between different types of software. In this work, while this assumption held for several different types of assessed software (i.e., database management, embedded OS, vehicle control), further verification of this relationship is required for all software. While ODC suggests that software defect classes are process and development agnostic, more evidence is required to justify this claim.

### 4. CASE STUDY

In this work, the ORCAS methodology was applied to an embedded smart sensor developed by VCU. The sensor is a barometric pressure and temperature sensing device that originates from the VCU Unmanned Aerial Vehicle (UAV) Laboratory [12]. The device consists of mature design and code, including the Software Requirements Specification (SRS) and Software Design Description (SDD) documentation, with over 10,000 hours of tested flight time. The software is written in GNU11 C programming language for the application code and runs on top of the ChibiOS Version 17.6.4 Real-Time Operating System (RTOS) [12]. The software was tested extensively using a pseudo-exhaustive test-based approach developed by VCU [10] and incorporates methodologies such as combinatorial testing, boundary value assessment, equivalence partitioning, and MCDC structural path coverage.

### 4.1. Pseudo Exhaustive Testing Results

The tests conducted by VCU were collected and assessed for testing completeness. The SRS and SDD documents were reviewed; and test tracing was conducted. In total, 10,687 tests were conducted on three functions: (1) 'circular_buffer_read;' (2) 'get_current_pressure;' and (3) 'kalman_filter.' Duplicate tests in T-way testing were not counted. Fault injection testing was completed by VCU; however, details on test formulation were not available, and thus, not counted. However, no new defects were detected through fault injection that were not originally caught by T-way combinatorial testing.

The defects that were discovered can be seen in Table 1. In column one and two, the testcase failure and root cause is described and conducted by VCU. In column 3, the defect class and resolution are specified via ORCAS.

**Table 1. Software failures, defect class, and resolution from VCU test data.**

| TestCase Failure | Root Cause | Defect Class and Resolution |
|---|---|---|
| Unable to fill the buffer completely. Can only fill buffersize-1 elements. | Incorrect buffer full check. | Algorithm defect, traversal method through circular buffer changed to check all elements. |
| TestExecution Timeout – Buffer overflow and corruption of neighboring memory addresses cause the 'Memcpy' function to hang when called with a length greater than the destination buffer size. | Missing destination buffer overflow check. | Checking defect, limit on size of buffer implemented via IF statement and truncation. |
| Indicates successful data read operation even with invalid configurations of buffer, 'size of buffer,' 'head,' and 'tail' pointers. | Invalid buffer configurations not handled. | Algorithm defect, changed true statement to false when invalid configuration branch taken. |
| Returns varying negative values of buffer read length when requested 'number of bytes' is negative. | Invalid negative values of the number of bytes to be read is not handled. | Checking defect, limit on negative inputs implemented via IF statement. |
| Negative values of buffer size are accepted during buffer initialization and buffer is filled with negative size value. | Invalid buffer size is not considered during buffer initialization. | Checking defect, limit on negative inputs implemented via IF statement. |
| Actual output value indicates 'Infinity.' | Missing divide by zero check. | Checking defect, try catch for divide by zero exception added. |
| Actual output value indicates 'NaN' (not a number). | Missing overflow check in float computation. | Checking defect, try catch for overflow exception added. |
| Function processes input values outside the valid range. | Missing invalid input value handling. | Checking defect, limit on inputs implemented via IF statement. |

## 4.2. Qualitative Evidence

The qualitative evidence collected in this case study include the RTM, as shown in
Table 2, the TCA, as shown in Table 3, and structural path coverage. The requirements were derived from the SRS document provided by VCU. In total, ten high-level requirements were identified. Additional requirements were also specified in the SDD but are refinements of the original ten. In column one, the requirement tag is provided, followed by a brief description in column two. In column three, tests that were implemented for each requirement were traced. If targeted tests for that requirement existed, a 'complete' grade (or 1) is assigned. Similarly, if no targeted tests were created, an 'incomplete' grade (or 0) is assigned. If tests that required part of the requirement existed, but were targeted at other requirements, an 'indirect' grade (or 0.5) is assigned. For example, the data averaging requirement (REQ-7) had direct tests associated and was completed with T-way combinatorial testing. For REQ-7 to be tested, collection of sensor data (REQ-1) must have been functional too. However, no tests were explicitly designed for REQ-1; therefore, only partial indirect testing was conducted on REQ-1 via testing of REQ-7. Of the ten high-level requirements, five had tests directly associated with them, four had indirect testing, and one had no tests.

For trigger coverage assessment, all three levels of software testing were assessed (Table 3). From right to left, the columns include the level of testing, the recommended activities, the triggers to be covered, VCU's implemented method for each activity, and the completeness score. Both component and subsystem testing were complete and had various types of tests associated. However, there were inadequate tests developed for the system level. Specifically, no tests were found for configuration or workload/stress testing. This corresponds with the RTM as no tests were traced to REQ-3. The startup/restart trigger also only had indirect tests (as the device had to be turned ON to run).

For the structural path coverage, VCU's team demonstrates that the use of T-way combinatorial testing can achieve 100% MCDC coverage [10]. Additional path assessment conducted via ORCAS agrees with VCU's results and revealed 12 unreachable but extraneous/benign code segments.

### 4.3. Quantitative Evidence

The quantitative evidence that was collected includes the eight defect reports from testing, as observed in Table 1, and the 10,687 tests. Each test was assumed to correspond to 1 hour of effort. In addition, reliability growth models could not be used for the VCU data as the timeline of effort required to create and run the tests was not measured. Therefore, the bounded failure probability estimation approach is adopted. The software failure modes are determined from the SRS and SDD documentation. Based on specifications, one of the hazards to the system was determined to be 'Incorrect pressure above $\pm 1\%$ of true value was provided to dependent devices causing unstable altitude adjustments.' The possible UIFs includes A, C, and D corresponding to a pressure reading that is missing when needed, a pressure reading that is asynchronous to reality, and a pressure reading that is too high or too low, NaN or Inf invalid values. UIF-B is not applicable for continuous monitoring systems as the value will always be needed.

Using the correlations shown in Figure 3, the individual UIF failure probabilities are determined, observed in Table 4. Defects with zero failure probability (or not detected) are excluded from the table. The total software failure probability was determined to be 5.854E-4 per hour (bottom right sum in Table 4). The probabilities of each UIF can be seen in the last row, which is the sum contribution of the individual probabilities.

**Table 2. Abridged Requirements Traceability Matrix**

| Specified Requirements | Functional Description | Test Complete | Scoring |
|---|---|---|---|
| Collection of Sensor Data (REQ-1) | ASCII format starting with six calibration constants followed by float point data. | Indirect | 0.5 |
| Transmission of Data (REQ-2) | I2C protocol transmitting [temp., pressure, KF-pressure]. | Indirect | 0.5 |
| Device Reconfiguration (REQ-3) | Capability at updating parameters of MS5611. | Incomplete | 0 |
| Re-ranging of Data (REQ-4) | Valid differential pressure (-1) to (+1) psi, valid absolute pressure range (0) to (15) psi, capability to re-scale pressures to defined ranges. | Complete | 1 |
| Temperature-effect Compensation (REQ-5) | Valid range (-40) to (125) ºC, capability to adjust Temp. to valid range. | Complete | 1 |
| Transmitter Calibration (REQ-6) | Recalibrate [min, max] of internal ranging and compensation parameters. | Complete | 1 |
| Data Averaging (REQ-7) | Analog data is converted using moving avg. Kalman filter with size of window updatable as user parameter. | Complete | 1 |
| Data Conversion (REQ-8) | Float to int conversion and rounding must be exist with error correction. | Complete | 1 |
| Data Output (REQ-9) | Manage serial transmission to host via UART. | Indirect | 0.5 |
| Data Logging/Clocking (REQ-10) | Host update rate must be greater than 2 Hz, with three commands to shell program. | Indirect | 0.5 |

### 4.4. ORCAS Results

The VCU smart sensor was designed to be a representation of a safety critical smart sensor device. As such, a thorough design, documentation, and development environment of the software was conducted. The device also had extensive operational hours justifying reliability. However, from the ORCAS assessment, we can conclude two specific things. From qualitative evidence, the inadequate areas of testing verification were identified. These primarily include the system level configurable options, stress, and communication with peripherals. In addition, while many conditions were tested, approximately 76%, this value may be unacceptable and used as an argument against software reliability. From the quantitative evidence, a maximum failure probability of 5.854E-4 per hour was determined based on the number of tests and number of detected defects. Note this value is for any software failure regardless of severity. While this value may seem acceptable, the qualitative evidence suggests that additional hidden defects may exist due to incomplete testing.

Table 3. Abridged Trigger Coverage Assessment

|  | Activity | Defect Triggers Required | Implemented Method | Score |
|---|---|---|---|---|
| **Component** | Unit Test | Simple Path | MCDC (Complete) | 1/1 |
|  | Function Test | Coverage, Variation, Sequence | T-way, BVA, EP, Sequence testing (Complete) | 3/3 |
| **Subsystem** | Unit Test | Simple Path, Complex Path | MCDC (Complete) | 2/2 |
|  | Function Test | Coverage, Variation, Sequence, Interaction | T-way, BVA, EP, Sequence testing, Interaction verification (Complete) | 4/4 |
| **System** | System Test | Startup/Restart | (Indirect) | 0.5/1 |
|  | System Test | Recovery/Exception Normal Mode | T-way, BVA, EP, fault injection (Complete) | 2/2 |
|  | System Test | Configuration Workload/Stress | (Incomplete) | 0/2 |

Table 4. Probabilities of each UIF with sum totals

|  | UIF – A | UIF – B | UIF – C | UIF – D | Total |
|---|---|---|---|---|---|
| **Algorithm** | 5.989E-5 | 0 | 6.550E-5 | 3.556E-5 | 1.609E-4 |
| **Checking** | 2.021E-4 | 0 | 1.437E-4 | 7.860E-5 | 4.244E-4 |
| **Total** | 2.620E-4 | 0 | 2.092E-4 | 1.142E-4 | 5.854E-4 |

## 5. CONCLUSION

A novel approach to determining software reliability and supporting evidence is discussed in this work. Software failure data from VCU's smart sensor device was collected and used to demonstrate the methodology. While a pseudo-exhaustive test-based approach was utilized (as recommended by ORCAS), it was shown that not all requirements and defect triggering scenarios were considered. These are areas where further testing effort are recommended. Software failure probabilities per UIF mode were also determined. A linear correlation between defect class and UCA/UIF was determined from 402 defect reports acquired from various open-source repositories. The individual and total UIF probabilities were determined from this correlation. By using ORCAS, the developer can identify areas where further work and the necessary methods are still needed to implement. They will also be able to provide evidence to stakeholders toward software reliability based on the qualitative and quantitative results. Future work includes collecting data from different types of software to further verify the UCA/UIF correlation developed in Figure 3.

**Acknowledgements**

The authors would like to thank Gordan E. Holt at Idaho National Laboratory for technical editing and formatting of this paper as well as Dr. Sai Zhang for technical review. The research activities and achievements documented in this paper were funded by the United States Department of Energy's Light Water Reactor Sustainability Program, Risk Informed Systems Analysis (RISA) Pathway. This submitted manuscript was authored by a contractor of the U.S. Government under DOE Contract No. DE-AC07-05ID14517. Accordingly, the U.S. Government retains and the publisher, by accepting the article for publication, acknowledges that the U.S. Government retains a nonexclusive, paid-up, irrevocable, worldwide license to publish or reproduce the published form of this manuscript, or allow others to do so, for U.S. Government purposes. This information was prepared as an account of work

**References**

[1] Nuclear Regulatory Commission, "Incorporation by Reference of Institute of Electrical and Electronics Engineerings Standard 603-2009, SECY-15-0106," Nuclear Regulatory Commission, Washington, 2016.

[2] Nuclear Regulatory Commission, "Guidance for Evaluation of Defense in Depth and Diversity to Address Common-Cause Failure due to Latent Design Defects in Digital Safety Systems, BTP 7-19," Nuclear Regulatory Commission, Washington, 2021.

[3] N. G. Leveson and J. P. Thomas, "STPA Handbook," MIT Partnership for Systems Approaches to Safety and Security, Cambridge, 2018.

[4] Electric Power Research Institute, "Hazard Analysis Methods for Digital Instrumentation and Control Systems - Revision 1," Electric Power Research Institute, Washington, 2021.

[5] A. D. Williams and A. J. Clark, "Using Systems Theoretic Perspectives for Risk-Informed Cyber Hazard Analysis in Nuclear Power Facilities," in *29th Annual INCOSE International Symposium*, Orlando, 2019.

[6] Electric Power Research Institute, "Digital Reliability Analysis Methodology," Electric Power Research Institute, Washington, 2021.

[7] International Electrotechnical Commission, "IEC 61508-1:2010," International Electrotechnical Commission, Geneva, 2010.

[8] J. D. Musa and K. Okumoto, "A Logarithmic Poisson Execution Time Model for Software Reliability Measurement," Institute of Electrical and Electronics Engineers, Whippany, 1984.

[9] H. S. Son, H. G. Kang and S. C. Chang, "Procedure for Application of Software Reliability Growth Models to NPP PSA," *Nuclear Engineering and Technology,* vol. 41, no. 8, pp. 1065-1072, 2009.

[10] A. Jayakumar, D. R. Kuhn, B. Simons, A. Collins, S. Gautham, R. Hite, R. N. Kacker, A. Rajagopala and C. Elks, "A Pseudo Exhaustive Software Testing Framework for Embedded Digital Devices in Nuclear Power," National Institute of Standards and Technology, Gaithersburg, 2021.

[11] G. D. Sirio, "ChibiOS/RT The Ultimate Guide," ChibiOS EmbeddedWare, 2020. [Online]. Available: https://www.chibios.org/dokuwiki/doku.php?id=chibios:documentation:books:rt:start. [Accessed 2022].

[12] C. Elks, C. Deloglos, A. Jayakumar, A. Tantawy, R. Hite and S. Guatham, "Specification of a Bounded Exhaustive Testing Study for a Software-based Embedded Digital Device," Idaho National Laboratory, Idaho, 2018.

[13] R. Chillarege, I. S. Bhandari, J. K. Chaar, M. J. Halliday, D. S. Moebus, B. K. Ray and M.-Y. Wong, "Orthogonal Defect Classification - A Concept for In-Process Measurements," *International Institute of Electrical Engineers Transactions on Software Engineering,* vol. 18, no. 11, pp. 943-956, 1992.

[14] International Business Machines, "Orthogonal Defect Classification v5.2 for Software Design and Code," International Business Machines, Armonk, 2013.

[15] M. Butcher, H. Munro and T. Kratschmer, "Improving software testing via ODC: Three Case Studies," *IBM Systems Journal,* vol. 41, no. 1, pp. 31-44, 2002.

[16] H. Bao, T. Shorthill and H. Zhang, "Redundancy-guided System-theoretic Hazard and Reliability Analysis of Safety-related Digital Instrumentation and Control Systems in Nuclear Power Plant," Idaho National Laboratory, Idaho, 2020.

[17] H. Bao, T. Shorthill, E. Chen and H. Zhang, "Quantitative Risk Analysis of High Safety-significant Safety-related Digital Instrumentation and Control Systems in Nuclear Power Plants using IRADIC Technology," Idaho National Laboratory, Idaho Falls, 2021.

[18] H. Bao, H. Zhang and K. Thomas, "An Integrated Risk Assessment Process for Digital Instrumentation and Control Upgrades of Nuclear Power Plants," Idaho National Laboratory, Idaho Falls, 2019.

[19] E. Chen, H. Bao, H. Zhang, T. Shorthill and N. Dinh, "Systems-theoretic hazard analysis of digital human-system interface relevant to reactor trip," in *12th Nuclear Plant Instrumentation, Control and Human-Machine Interface Technologies*, 2021.

[20] H. Bao, H. Zhang, T. Shorthill and S. Lawrence, "Quantitative Evaluation of Common Cause Failures in High Safety-significant Safety-related Digital Instrumentation and Control Systems in Nuclear Power Plants," 7 April 2022. [Online]. Available: https://arxiv.org/abs/2204.03717. [Accessed 2022].

[21] H. Zhang, H. Bao, T. Shorthill and E. Quinn, "An Integrated Risk Assessment Process of Safety-Related Digital I&C Systems in Nuclear Power Plants," 17 Dec 2021. [Online]. Available: arXiv preprint arXiv:2112.09287. [Accessed 2022].

[22] T. Shorthill, H. Bao, H. Zhang and H. Ban, "A redundancy-guided approach for the hazard analysis of digital instrumentation and control systems in advanced nuclear power plants," *Nuclear Technology,* vol. 208, no. 5, pp. 892-911, 2021.

[23] H. Bao, T. Shorthill and H. Zhang, "Hazard analysis for identifying common cause failures of digital safety systems using a redundancy-guided systems theoretic approach," *Annals of Nuclear Energy,* vol. 148, no. 1, p. 107686, 2020.

[24] R. N. K. Y. L. Richard D. Kuhn, "Practical Combinatorial Testing," National Institute of Standard Technology, 2010.

[25] K. J. Hayhurst, D. S. Veerhusen, J. J. Chilenski and L. K. Rierson, "A Practical Tutorial on Modified Condition / Decision Coverage," National Aeronautics and Space Administration, Washington, 2001.

[26] S. C. Reid and S. Shrivenham, "An Empirical Analysis of Equivalence Partitioning, Boundary Value Analysis and Random Testing," in *Proceedings Fourth International Software Metrics Symposium*, Como, 1997.

[27] M. Li and C. S. Smidts, "A Ranking of Software Engineering Measures Based on Expert Opinion," *Institute of Electrical and Electronics Engineers Transactions on Software Engineering,* vol. 29, no. 9, pp. 811-824, 2003.

[28] R. Chillarege, "Orthoginal Defect Classification," in *Handbook of Software Reliability Engineering*, McGraw-Hill Book Company, 1996, pp. 359-399.

[29] J. Agnelo, N. Laranjeiro and J. Bernardino, "Using Orthoginal Defect Classification to Characterize NoSQL Database Defects," *The Journal of Systems and Software,* vol. 159, 2020.

[30] M. C. Kim, S. C. Jang and J. Ha, "Possibilities and Limitations of Applying Software Reliability Growth Models to Safety-Critical Software," *Nuclear Engineering and Technology,* vol. 39, no. 2, pp. 129-132, 2007.

[31] A. Wood, "Software Reliability Growth Models," Hewlett-Packard, Palo Alto, 1996.

[32] T. Shorthill, H. Bao, H. Zhang and H. Ban, "A Novel Approach for Software Reliability Analysis of Digital Instrumentation and Control Systems in Nuclear Power Plants," *Annals of Nuclear Energy,* vol. 158, 2021.